\documentclass[10pt,final, twocolumn]{IEEEtran}
\usepackage{lipsum}
\usepackage[utf8]{inputenc}
\usepackage{amsmath,amsfonts,amssymb,bm}
\usepackage{graphicx}
\usepackage{subcaption}
\usepackage{xcolor}
\usepackage{mathtools}
\usepackage{comment}
\usepackage{textcomp}

\usepackage{multirow, flushend}

\newcommand{\bi}{\begin{itemize}}
\newcommand{\ei}{\end{itemize}}

\newcommand{\be}{\begin{enumerate}}
\newcommand{\ee}{\end{enumerate}}
\newcommand{\bd}{\begin{description}}
\newcommand{\ed}{\end{description}}
\newcommand{\bc}{\begin{center}}
\newcommand{\ec}{\end{center}}
\newcommand{\bt}{\begin{tabbing}}
\newcommand{\et}{\end{tabbing}}
\newcommand{\bfig}{\begin{figure}}
\newcommand{\efig}{\end{figure}}
\newcommand{\beq}{\begin{equation}}
\newcommand{\beqarr}{\begin{eqnarray}}
\newcommand{\beqarrn}{\begin{eqnarray*}}
\newcommand{\eeq}{\end{equation}}
\newcommand{\eeqarr}{\end{eqnarray}}
\newcommand{\eeqarrn}{\end{eqnarray*}}
\newcommand{\bflr}{\begin{flushright}\vspace{-0.2in}}
\newcommand{\eflr}{\end{flushright}}
\newcommand{\bsub}{\begin{subequations}}
\newcommand{\esub}{\end{subequations}}
\newcommand{\barr}{\begin{array}}
\newcommand{\earr}{\end{array}}
\newcommand{\nn}{\nonumber}

\def\undb#1{\mbox{\bf{#1}}}

\def\BibTeX{{\rm B\kern-.05em{\sc i\kern-.025em b}\kern-.08em
		T\kern-.1667em\lower.7ex\hbox{E}\kern-.125emX}}

\begin{document}

\title{\huge{RIS-Assisted MIMO CV-QKD at THz Frequencies: Channel Estimation and SKR Analysis}}
\author{Sushil Kumar, Soumya~P.~Dash,~\IEEEmembership{Senior Member,~IEEE,} Debasish~Ghose, \IEEEmembership{Senior Member, IEEE,}\\ and
George C. Alexandropoulos, \IEEEmembership{Senior Member, IEEE}

\thanks{S. Kumar and S. P. Dash are with the School of Electrical Sciences, Indian Institute of Technology Bhubaneswar, Argul, Khordha, 752050 India e-mail: (a24ec09010@iitbbs.ac.in, soumyapdashiitbbs@gmail.com).}
\thanks{D. Ghose is with the School of Economics, Innovation, and Technology at Kristiania University College, Bergen (e-mail: Debasish.Ghose@kristiania.no).}
\thanks{G. C. Alexandropoulos is with the Department of Informatics and Telecommunications, National and Kapodistrian University of Athens, Panepistimiopolis Ilissia, 15784 Athens, Greece and also with the Department of Electrical and Computer Engineering, University of Illinois Chicago, IL 60601, USA (e-mail: alexandg@di.uoa.gr).}
}
\maketitle

\begin{abstract}
In this paper, a multiple-input multiple-output (MIMO) wireless system incorporating a reconfigurable intelligent surface (RIS) to efficiently operate at terahertz (THz) frequencies is considered. The transmitter, Alice, employs continuous-variable quantum key distribution (CV-QKD) to communicate secret keys to the receiver, Bob, which utilizes either homodyne or heterodyne detection. The latter node applies the least-squared approach to estimate the effective MIMO channel gain matrix prior to receiving the secret key, and this estimation is made available to Alice via an error-free feedback channel. An eavesdropper, Eve, is assumed to employ a collective Gaussian entanglement attack on the feedback channel to avail the estimated channel state information. We present a novel closed-form expression for the secret key rate (SKR) performance of the proposed RIS-assisted THz CV-QKD system. The effect of various system parameters, such as the number of RIS elements and their phase configurations, the channel estimation error, and the detector noise, on the SKR performance are studied via numerical evaluation of the derived formula. It is demonstrated that the RIS contributes to larger SKR for larger link distances, and that heterodyne detection is preferable over homodyne at lower
pilot symbol powers.  
\end{abstract}
\section{Introduction}
The upcoming sixth-generation (6G) of wireless networks is expected to attain extraordinary capabilities, featuring peak data rates in the order of Tbps, ultra-low latency, and a twofold increase in both spectrum and energy efficiencies relative to its preceding fifth-generation (5G)~\cite{9669056, GiPoMe:cm20}. 
To satisfy these stringent requirements, various supporting technologies are being investigated, including ultra massive multiple-input multiple-output (MIMO) systems \cite{9825647, 9585108,XYA2024}, reconfigurable intelligent surfaces (RISs) \cite{HuangTWC2019,RDP2019, 10256045, 10463684, 10200235}, and the utilization of terahertz (THz) frequency ranges \cite{9766110, 9546670}.

As network capacity and connectivity increase, security issues escalate, requiring innovative strategies for secure communication links. Quantum key distribution (QKD) is emerging as a viable approach for ensuring robust, quantum-resistant security and vulnerability to eavesdropping. 
The significant advancements in quantum computing make traditional higher-layer encryption schemes based on the Rivest-Shamir-Adleman (RSA) algorithm susceptible to Shor's factoring algorithm \cite{weedbrook2012gaussian, manzalini2020quantum}. Similarly, the security of physical-layer encryption, which uses conventional key distribution algorithms like Diffie-Hellman, is insecure because it relies on the assumption that classical computers cannot solve the computationally intensive problem of distinct logarithms in a reasonable amount of time~\cite{diffie1976new}. Consequently, contemporary computationally secure encryption techniques may be compromised due to the significant advancements in practical quantum computing. QKD enables safe transmissions of secret keys between two entities, such as Alice and Bob, which can be utilized for one-time-pad cryptography for 6G the physical-layer applications~\cite{neel_spd_channel_estimation_skr}. QKD uses quantum mechanics to safely send cryptographic keys, providing proven protection from the threats that come with quantum computing~\cite{8865103, 9845436, 9552894}.
There are two principal categories of QKD protocols~\cite{8865103, cv_dv_qkd_Scarani_2009}: discrete variable QKD (DV-QKD), which utilizes single-photon sources along with polarization or time-bin encoding, and continuous variable QKD (CV-QKD), which employs conventional coherent optical sources and homodyne or heterodyne detectors, facilitating integration with existing telecommunications infrastructure~\cite{weedbrook2012gaussian, 8732438}. DV-QKD offers strong security, but its reliance on complicated single-photon technology makes it difficult to use on a large scale~\cite{9102386, 9799745}. On the other hand, CV-QKD is easier to integrate with the existing hardware for wireless communications, opening the door to scalable, quantum-secure 6G networks~\cite{lodewyck2007quantum, cvqkd_ppf_2024}.

On another front, RISs have recently emerged as a promising technology for controlling the propagation of electromagnetic waves in wireless systems~\cite{JAB2022,10255749,ASD2021}. By adjusting the phase of incoming signals, an RIS can manipulate the properties of reflected waves, thereby improving signal coverage, strength, and reliability \cite{BAL2024,10385147,RIS_najafi9443170}, and thus, it is aimed to operate across a broad range of frequencies, including microwave and millimeter-wave (mmWave)~\cite{SAW2021a}, as well as THz bands~\cite{chen_ris_thz21_9690477,RIS_TH9840765,MML2024}. Specifically, RISs have been shown to alleviate the challenges in wireless channels faced due to obstructions, signal degradation, or restricted line-of-sight (LoS) pathways~\cite{APK2023}, especially at elevated frequencies like mmWave and THz~\cite{AJG2024}, resulting in communication systems with ultra-high speeds, minimal latency, and extensive connectivity \cite{ASA2024, 9698029, 9802114, 10400440, 10388479, 10287142, risFSo_21_Jamali_Najafi_9627820}.

Over the past few years, researchers have attempted to combine the latter emerging technologies to examine the feasibility of THz CV-QKD systems. The authors in~\cite{wang2019inter} studied quantum communications with QKD within a micro-satellite framework in low-earth-orbit scenarios. A physical hardware architecture of a QKD system operating at the THz was designed in~\cite{ottaviani2020terahertz}. The viability of elevated key rates and extended distances in THz CV-QKD by applying multi-carrier multiplexing in inter-satellite communication links was demonstrated in~\cite{liu2021multicarrier}. However, in all these studies, the presented THz QKD systems attained low secret key rates (SKRs) and short maximum transmission distances due to the high atmospheric absorption loss and free-space path loss. To overcome this, the use of MIMO was found to be a viable option. The authors in~\cite{2_ref_paper} considered a MIMO THz QKD scheme capable to obtain positive SKRs in the $10-30$ THz frequency spectrum. The impact of the channel estimation error on SKR for MIMO CV-QKD wireless communication system at THz was studied in~\cite{neel_spd_channel_estimation_skr}.  

Very recently, the technology of RISs was considered in QKD system designs. The authors in \cite{neel_IRS_assisted_nlosqkd_10289124} analyzed the SKR performance under log-normal fading turbulence conditions for an RIS-assisted free-space optical system. An RIS-assisted MIMO THz CV-QKD system that attains a high SKR and extensive transmission lengths compared to a non-RIS counterpart was presented in~\cite{sushil2024risassistedthzmimo}. However, both studies assumed the possession of flawless channel information by both Alice and Bob, which is impractical. In fact, to the best of the authors' knowledge, there does not exist any investigation in the open technical literature of the impact of channel estimate errors on the SKR of RIS-assisted MIMO CV-QKD systems.

Motivated by the latter research gap, in this paper, we consider an RIS-assisted MIMO THz system where the transmitter, Alice, wishes to send secret keys to the receiver, Bob, using CV-QKD. A pragmatic channel estimation approach, based on the least-squares (LS) technique, is employed at Bob to estimate the effective MIMO channel prior to secret key reception. The estimated channel state information is then provided to Alice via an error-free feedback channel, which is attacked by an eavesdropper, Eve, using a collective Gaussian entanglement attack. We analyze the SKR performance of the system considering the impact of channel estimate errors into the transmit-receive model during the key generation phase. The contributions of the paper are summarized as follows:
\begin{itemize}
    \item We present an LS-based channel estimation protocol according to which Alice and Bob design their beamforming matrices via the singular-value decomposition (SVD) of the estimated effective MIMO channel matrix.
    \item The input-output relationship between Alice and Bob throughout the key generation phase is established, which includes the supplementary noise factors resulting from channel estimate inaccuracies as well as the detector noise arising due to Bob employing either homodyne or heterodyne detection for the secret key reception.
    \item The variance of the latter supplementary noise components is delineated to obtain a novel closed-form expression for the SKR performance of the considered RIS-assisted MIMO CV-QKD system. The derived expression considers both types of measurements at Bob's side, integrating the impacts of channel estimation overhead as well as the supplementary noise components resulting from imperfect channel estimation.
    \item Comprehensive numerical simulations are conducted to investigate the impact of the channel estimate error and evaluate the influence of critical parameters, such as pilot length and pilot power, on the system's SKR performance. The roles of the number and phase configuration of the RIS elements on the SKR metric are examined, as well as those of the choice of the measurement approach at Bob and the physical distance between Alice and Bob.
\end{itemize}

The rest of the paper is organized as follows. The RIS-assisted MIMO CV-QKD system model and the THz channel model are presented in Section~II. The proposed channel estimation protocol is introduced in Section~III. Section IV details the secret key generation, transmission, and measurement phases based on the evaluation of the covariance matrix of the noise arising due to imperfect channel information. The SKR analysis for both homodyne and heterodyne measurements considering that Eve deploys a collective Gaussian entanglement attack is presented in Section~V. The numerical results of the paper are presented in Section~VI, followed by its concluding remarks in Section~VII.

\textit{Notation:} Bold $(\textbf{A})$ and $(\boldsymbol{a})$ letters stand for matrices and vectors, respectively. $\textbf{A}^\dagger$ symbol stands for the conjugate transpose of $\textbf{A}$, while $\textbf{A}^T$ its transpose. Notations $\boldsymbol{1}_{M\times N}$ and $\boldsymbol{0}_{M\times N}$ represent a $M \times N$ matrix consisting of all ones and all zeros, respectively, and $\jmath\triangleq\sqrt{-1}$. $\textbf{I}_M$ represents the $M \times M$ identity matrix, and $\text{diag}(\boldsymbol{a})$ with $\boldsymbol{a} \in \mathbb C^M$ returns a $M \times M$ diagonal matrix with the elements of $\boldsymbol{a}$ on its principal diagonal. $a^*$ represents the conjugate of $a$. Notation $\mathcal{N} (\boldsymbol{\mu},\boldsymbol{\Sigma})$ represents the real multivariate Gaussian distribution, where $\boldsymbol{\mu}$ is the mean vector and $\boldsymbol{\Sigma}$ is the covariance matrix, $|\cdot|$ denotes the magnitude operator. Finally, $\undb{E}\left[\cdot\right]$ is the expectation operator and $\text{Tr} (\cdot )$ is the trace operator.
\section{System Model}
\begin{figure}[ht]
    \centering
    \begin{subfigure}{\columnwidth}
        \centering
        \includegraphics[height=2.1in,width=2.6in]{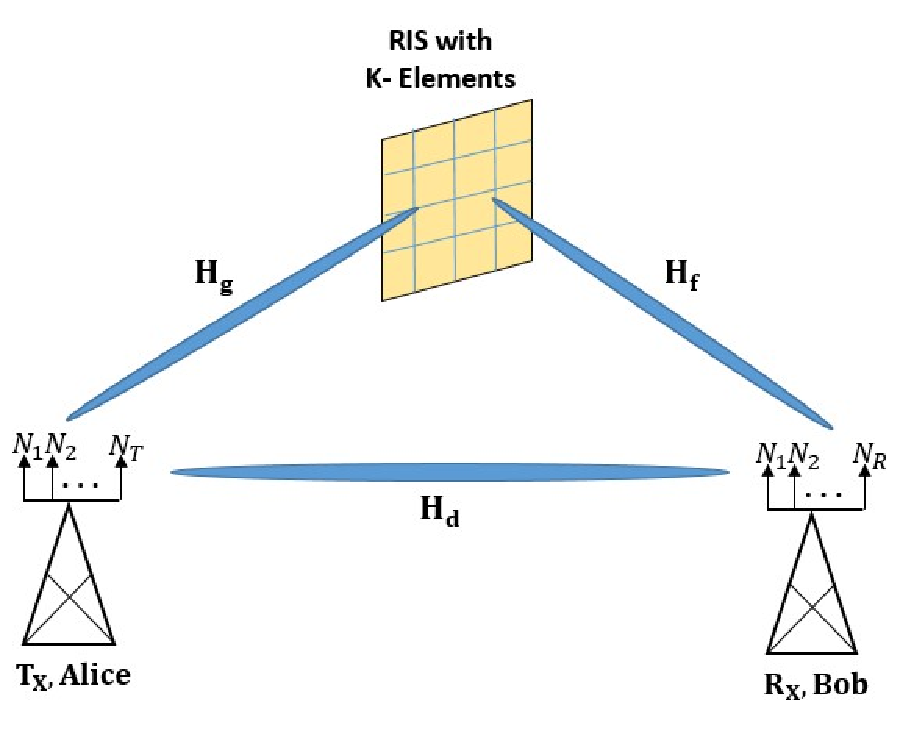}
        \caption{}
        \label{fig;1a}
    \end{subfigure}
    \hfill
    \begin{subfigure}{\columnwidth}
        \centering
        \includegraphics[height=2.1in,width=3in]{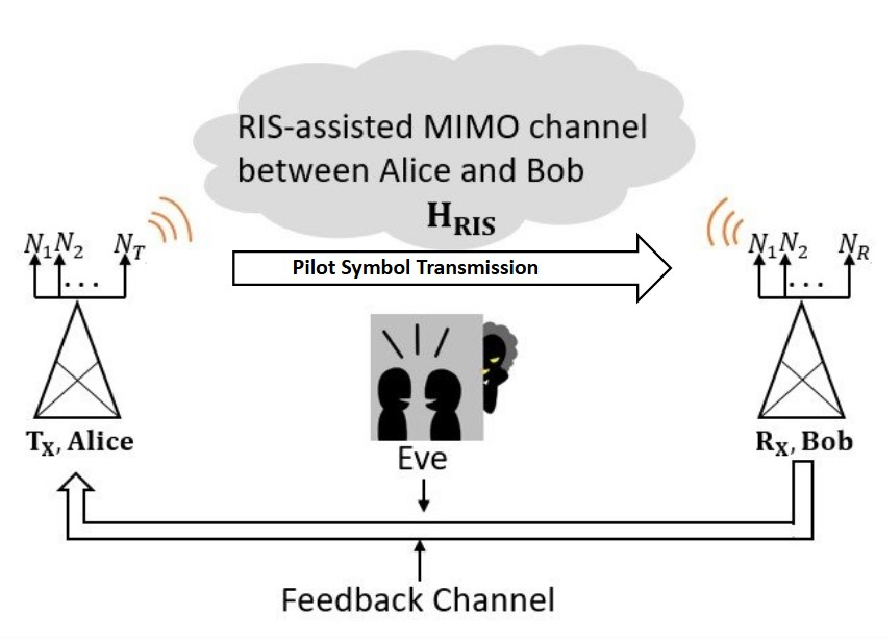}
        \caption{}
        \label{fig;1b}
    \end{subfigure}
    \caption{(a) The considered RIS-assisted MIMO CV-QKD wireless communication system at THz frequencies; (b) The feedback channel designed to transfer the estimated MIMO channel at Bob's side to Alice, which is eavesdropped by Eve.}
    \label{fig:1}
    \vspace{-1cm}
\end{figure}

We consider an RIS-assisted MIMO CV-QKD setup as depicted in Fig.~\ref{fig;1a}. The transmitter, referred to as Alice, and the receiver, referred to as Bob, are equipped with $N_{T}$ and $N_{R}$ antennas, respectively, and there exists a direct LoS channel between them together with a virtual LoS channel via an RIS.
The RIS comprises $K$ passive reflecting elements that can be optimized to adjust the phases of the incoming information-bearing signals~\cite{ASD2021}. Thus, the overall MIMO channel between Alice and Bob considering the RIS-assisted path as well as the direct Alice-Bob path, which is denoted by $\textbf{H}_{\text{RIS}}\in \mathbb{C}^{N_{R}\times N_{T}}$, can be expressed as follows:
\beq
\textbf{H}_{\text{RIS}} \triangleq \textbf{H}_{d} + \textbf{H}_{f} \mathbf{\Phi}_{\text{RIS}} \textbf{H}_{g},
\label{eq1}
\eeq
where $\mathbf{\Phi}_{\text{RIS}} \triangleq \text{diag} \left(e^{\jmath \phi_{1}}, \ldots,e^{\jmath \phi_{K}}\right)$ with $\phi_k$ denoting the phase shift introduced by each $k$-th ($k=1,\ldots,K$) reflective element of the RIS. Furthermore, $\textbf{H}_{d} \in \mathbb{C}^{N_{R}\times N_{T}}$, $\textbf{H}_{g} \in \mathbb{C}^{K\times N_{T}}$, and $\textbf{H}_{f}\in \mathbb{C}^{N_{R} \times K}$ are the channel gain matrices between Alice and Bob, Alice and RIS, and RIS and Bob, respectively, which are modeled for the considered in this paper THz frequency spectrum as follows:

\beqarr
&& \! \! \! \! \! \! \! \! \! \!
\! \! \! \! \! \! \! \! \! \! \textbf{H}_{d} \triangleq
\sum_{l=1}^L \sqrt{\delta_{d,l}}
e^{\jmath 2 \pi f_c \tau_{d,l}} 
\textbf{h}_{N_{R}}\left( \theta_l^{R_X} \right)\textbf{h}^\dagger_{N_{T}}
\left(\theta_{l}^{T_X} \right), \nn \\
&& \! \! \! \! \! \! \! \! \! \!
\! \! \! \! \! \! \! \! \! \! \textbf{H}_{g} \triangleq
\sum_{m=1}^M \sqrt{\delta_{g,m}}
e^{\jmath 2 \pi f_c \tau_{g,m}} \textbf{h}_{\text{RIS}} 
\left(\varphi,\theta_m^{\text{RIS}}\right)
\textbf{h}^\dagger_{N_{T}}
\left( \theta_m^{T_X} \right), \nn \\
&& \! \! \! \! \! \! \! \! \! \!
\! \! \! \! \! \! \! \! \! \! \textbf{H}_{f} \triangleq
\sum_{n=1}^N \sqrt{\delta_{f,n}}
e^{\jmath 2 \pi f_c \tau_{f,n}}
\textbf{h}_{N_{R}} \left( \theta_n^{R_X} \right) 
\textbf{h}^\dagger_{\text{RIS}}
\left( \varphi,\theta_n^{\text{RIS}} \right),
\label{eq2}
\eeqarr
where $L$, $M$, and $N$ denote the number of signal propagation paths within the wireless channels $\textbf{H}_d$, $\textbf{H}_g$, and $\textbf{H}_f$, respectively. In addition, $f_c$ is the carrier frequency, and $\tau_{d,l}$, $\tau_{g,m}$, $\tau_{f,n}$ and $\delta_{d,l}$, $\delta_{g,m}$, $\delta_{f,n}$ are the propagation delays and path losses corresponding to the $l$-th path of  $\textbf{H}_d$, $m$-th path of $\textbf{H}_g$, and $n$-th path of $\textbf{H}_f$, respectively. Moreover, $\theta_l^{T_X}$,$\theta_m^{T_X}$, and $\theta_n^{\text{RIS}}$ represent the respective angles of departure (AoDs), and $\theta_l^{R_X}$, $\theta_m^{\text{RIS}}$, and $\theta_n^{R_X}$ are the respective angles of arrival (AoAs) from the transmitter's and to the receiver's uniform linear arrays (ULAs), respectively. The antenna elements in both ULAs are considered to be uniformly placed in a single dimension such that the inter-element spacing is maintained at $d_a$. This results in the following expression for both the array response vectors $\undb{h}_{N_{T}}$ and $\undb{h}_{N_{R}}$:
\beq
\undb{h}_{N} \left( \theta \right) \triangleq \frac{1}{\sqrt{N}} 
\left[1, e^{\jmath \frac{2\pi}{\lambda_c}
d_a \sin \theta} , \ldots ,
e^{\jmath \frac{2\pi}{\lambda_c}
d_a \left( N-1 \right) \sin \theta} \right]^T,
\label{eq3}
\eeq
where $N \in \{N_{T},N_{R}\}$ and $\lambda_c = c/f_c$ with $c$ being the speed of light. Finally, the vector $\textbf{h}_{\text{RIS}}\in \mathbb C^{K\times 1}$ in (\ref{eq2}) indicates the ULA of the RIS and is given as follows~\cite{path_los_RIS_Ellingson_2021,ris_assisted_mmwave_mimo_9918631}:
\beqarr
&& \! \! \! \! \! \! \! \! \!
\! \! \! \! \! \! \! \! \! \!
\textbf{h}_{\text{RIS}} 
\left( \varphi, \theta_{i}^{\text{RIS}} \right)
\triangleq \frac{1}{\sqrt{K}} \left[e^{\jmath\frac{2\pi}{\lambda_c}
\left(\vartheta_X^{\varphi, \theta_{i}^{\text{RIS}}} 
+ \vartheta_Y^{\varphi, \theta_{i}^{\text{RIS}}} \right)}, \cdots \right. \nn \\
&& \qquad \left. \ldots ,
e^{\jmath\frac{2\pi}{\lambda_c}
\left( \left( K_X-1 \right) \vartheta_X^{\varphi, \theta_{i}^{\text{RIS}}}
+ \left( K_Y - 1 \right) \vartheta_Y^{\varphi, \theta_{i}^{\text{RIS}}} \right)} \right]^T,
\label{eq4}
\eeqarr
where we have used the definitions for $i=m$ and $i=n$:
\beqarr 
\vartheta_X^{\varphi, \theta_{i}^{\text{RIS}}}
\! \! \! \! &\triangleq& \! \! \! \! d_X \cos\left(\varphi\right)
\sin\left(\theta_{i}^{\text{RIS}}\right),\nn\\
\vartheta_Y^{\varphi, \theta_{i}^{\text{RIS}}}
\! \! \! \! &\triangleq& \! \! \! \! d_Y \sin\left(\varphi\right)
\sin \left(\theta_i^{\text{RIS}} \right).
\label{eq5}
\eeqarr
The RIS elements are considered to be arranged along a two-dimensional structure~\cite{ris_review_2023, ris_channel} with $K_X$ and $K_Y$ reflecting elements being placed along the horizontal and vertical axes, respectively, implying that $K_X K_Y=K$. Further, the separation between the elements in the corresponding axes is denoted by $d_X$ and $d_Y$, and the path losses in (\ref{eq2}) are expressed as:
\beq
\delta_{j,i} \triangleq \!
\begin{cases} \! \! 
    \left( \frac{\lambda_c}{4\pi d_{j,i}} \right)^2 
    \! \! G_T G_R 10^{-0.1 \rho d_{j,i}} \, , j \in \left\{d, g, f \right\} , i=1 \\
    \! \varsigma \xi_i \left( \frac{\lambda_c}
    {4\pi d_{j,i}} \right)^2 \! \! G_T G_R
    10^{-0.1 \rho d_{j,i}}
    \! , i \in \left\{l, m, n \neq 1 \right\}
\end{cases}\!\!\!\!\!\!,
\label{eq6}
\eeq
where $d_{j, i}$ is the smallest path length, $\xi_i$ is the Fresnel reflection coefficient of each $i$-th multipath component, $\varsigma$ is the Rayleigh roughness factor, and $\rho$ (in dB/km) is the atmospheric absorption loss. Moreover, the ULA gains at Alice and Bob are $G_T \triangleq N_{T}G_a$ and $G_R \triangleq N_{R}G_a$, respectively, where $G_a$ represents the gain of each component of the transmitter and receiver antennas and $G_T$ (or $G_R) = K$ when the data symbols are transmitter from (or to) the RIS, respectively.
\section{Proposed LS Channel Estimation Protocol} 
In this paper, we focus on the practical setting for our RIS-assisted MIMO CV-QKD system where Bob performs estimation of the effective MIMO channel $\textbf{H}_{\text{RIS}}$ in (\ref{eq1}) before the actual key distribution protocol between this receiver and Alice becomes operational. As previously discussed and depicted in Fig.~\ref{fig;1a}, our system supports a seamless feedback link between Bob and Alice, where the former communicates to the latter its estimated channel parameters that will be used in the next phase by Alice for the key distribution and the actual data transmission. We assume that the eavesdropper Eve attempts to intercept the feedback link to obtain Bob's estimation, i.e., the estimation of the RIS-assisted MIMO between Alice and Bob, as shown in Fig.~\ref{fig;1b}.

To enable channel estimation, Alice transmits $L_p$ pilot signals to Bob via the effective MIMO channel $\undb{H}_{\text{RIS}}$~\cite{JAS2022}. It is assumed that this channel is quasi-static, implying that $\mathbf{H_{\text{RIS}}}$ remains constant throughout the channel coherence time $T_c$, thus, it must hold $L_p < T_c$. The pilot symbols are constructed by using the position and momentum quadratures to create Gaussian coherent states~\cite{2_ref_paper, QC_book}, hence, the $N_{T} \times 1$ pilot symbol vector during each $\ell$-th transmission duration ($\ell=1,\ldots, L_p$) is represented by $\boldsymbol{\psi}_p^\ell \triangleq \mathbf{x}_p^\ell +\jmath \mathbf{p}_p^\ell$, where $\mathbf{x}_p^\ell \triangleq [x_{p,1}^\ell, \dots, x_{p,N_{T}}^\ell]^T$, $\mathbf{p}_p^\ell \triangleq [p_{p,1}^\ell, \dots, p_{p,N_{T}}^\ell]^T$, and $\boldsymbol{\psi}_p^\ell \sim {\mathcal{CN}} \left( \undb{0}_{N_{T}} , 2 V_p \undb{I}_{N_{T}} \right)$. Upon reception of the pilot symbols, Bob performs heterodyne measurements, implying that both the position and momentum quadratures of the received signal mode are measured. To this end, the complex-valued $N_{R}$-element received vector at each $\ell$-th pilot duration can be mathematically expressed as follows:
\beq
\mathbf{y}_p^\ell \triangleq \mathbf{H}_{\text{RIS}} \boldsymbol{\psi}_p^\ell
+ \mathbf{H}_{\text{RIS}} \boldsymbol{\psi}_0^\ell
+ \mathbf{n}_{\text{het}}^\ell,
\label{eq7}
\eeq
where $\mathbf{n}_{\text{het}}^\ell \triangleq \mathbf{n}_{\text{het},I}^\ell + \jmath \mathbf{n}_{\text{het},Q}^\ell$ is the additive noise vector at the receiver arising due to the heterodyne measurement, with $\mathbf{n}_{\text{het}, I}^\ell, \mathbf{n}_{\text{het}, Q}^\ell \sim \mathcal{N}\left(\mathbf{0}_{N_{R} \times 1}, \left( 2\nu_{en} + 1\right) \undb{I}_{N_{R}} \right)$ being respectively the in-phase and quadrature-phase components, respectively, where $\nu_{en}$ being is variance of the electronic noise. Note that, $\boldsymbol{\psi}_0^\ell = \mathbf{x}_0^\ell +\jmath \mathbf{p}_0^\ell$ is the preparation thermal noise at the side of Alice with $\mathbf{x}_0^\ell, \mathbf{p}_0^\ell \sim \mathcal{N}\left(\mathbf{0}_{N_{T} \times 1}, V_0 \undb{I}_{N_{R}} \right)$. Here $V_0\triangleq2\Bar{n}+1$ is the thermal noise variance with $\Bar{n}\triangleq[\exp\left(hf_c/k_BT_e\right)-1]^{-1}$, where $h$ is the Plank's constant, $k_B$ is the Boltzmann's constant, and $T_e$ is the environmental temperature in Kelvin. By collecting all $L_p$ pilot symbols, the following vector-matrix expression is deduced from~(\ref{eq7}):
\beq
\mathbf{Y}_p = \mathbf{H}_{\text{RIS}} \boldsymbol{\Psi}_p + \mathbf{H}_{\text{RIS}} \boldsymbol{\Psi}_0 + \mathbf{N}_{\text{het}}, 
\label{eq8}
\eeq
where $\mathbf{Y}_p \triangleq [\mathbf{y}_p^1, \ldots, \mathbf{y}_p^{L_p}] \in \mathbb{C}^{N_{R} \times L_p}$, $\mathbf{\Psi}_p \triangleq [\mathbf{x}_p^1, \ldots, \mathbf{x}_p^{L_p} ] \in \mathbb{C}^{N_{T} \times L_p}$, and $\mathbf{N}_{\text{het}} \triangleq [\mathbf{n}_{\text{het}}^1, \ldots, \mathbf{n}_{\text{het}}^{L_p}] \in \mathbb{C}^{N_{R} \times L_p}$. By defining $\mathbf{\tilde{N}} \triangleq \mathbf{H}_{\text{RIS}} \boldsymbol{\Psi}_0 + \mathbf{N}_{\text{het}}$ as the equivalent noise matrix, (\ref{eq8}) can be re-written as follows:
\beq
\mathbf{Y}_p = \mathbf{H}_{\text{RIS}} \boldsymbol{\Psi}_p + \mathbf{\tilde{N}}.
\label{eq9}
\eeq
As expected, to enable explicit channel estimation, the pilot matrix $\boldsymbol{\Psi}_p$ is known at both Alice and Bob. However, the covariance matrix of $\mathbf{\tilde{N}}$ included in~(\ref{eq8}), including the unknown $\mathbf{H}_{\text{RIS}}$, is unknown. To this end, by applying the LS approach, $\mathbf{H}_{\text{RIS}}$ can be estimated from~(\ref{eq8}) at Bob as follows:
\beqarr
\mathbf{H}_{\text{RIS}_{\text{LS}}} \! \! \! \! &\stackrel{(a)}{=}& \! \! \! \! \mathbf{Y}_p \boldsymbol{\Psi}_p^\dagger\left(\boldsymbol{\Psi}_p\boldsymbol{\Psi}_p^\dagger \right)^{-1} \nn \\
&\stackrel{(b)}{=}& \! \! \! \! \mathbf{H}_{\text{RIS}} + \underbrace{ \mathbf{ \tilde{N}} \boldsymbol{ \Psi}_p^ \dagger \left( \boldsymbol{ \Psi}_p \boldsymbol{ \Psi}_p^ \dagger \right)^{-1}}_{\triangleq\Delta \mathbf{H}_{\text{RIS}}} \, ,
\label{eq10}
\eeqarr
where step $(b)$ is obtained by substituting (\ref{eq9}) in $(a)$. Note that LS-optimal channel estimation performance is obtained when $\boldsymbol{\Psi}_p$ is constructed such that its rows are orthogonal and the norm of each row is equal to $\sqrt{V_p L_p}$. These conditions are met when using the discrete Fourier transform (DFT) matrix \cite{neel_spd_channel_estimation_skr}. In this case, $\boldsymbol{\Psi}_p$ is given by:  
\beqarr
\! \! \! \! \! \! \! \! \! \! \! \! \!
\! \! \! \! \! \! \! \! \! \! \! \! \!
&& \mathbf{\Psi_p} = \sqrt{V_p} \nn \\
\! \! \! \! \! \! \! \! \! \! \! \!
\! \! \! \! \! \! \! \! \!
&& \times \begin{bmatrix}
1  & 1 & \ldots & 1 \\
1 & e^{\jmath2\pi/L_p} & \ldots & e^{\jmath2\pi \left(L_p-1 \right)/L_p} \\
\vdots & \vdots & \ldots & \vdots \\
1 & e^{\jmath2\pi \left(N_{T}-1 \right)/L_p} & \ldots & e^{\jmath2\pi \left(N_{T}-1 \right) \left(L_p-1 \right)/L_p} \\
\end{bmatrix} \! .
\label{eq11}
\eeqarr
When the estimated channel matrix is obtained, Bob occupies the feedback channel, which is assumed to be a classical public authenticated channel, to transfer the estimated coefficients of the RIS-assisted MIMO channel to Alice.

\section{Secret Key Generation and Processing}
After acquiring the estimated channel, Alice employs CV-QKD based on Gaussian modulation to encrypt the secret keys intended for Bob via the RIS-assisted wireless channel. To attain this, the two $N_T$-element random vectors $\mathbf{x}_{\text{A}}$ and $\mathbf{p}_{\text{A}}$ are generated, which correspond to the position and momentum quadratures of the secret key states. These vectors are statistically independent and adhere to zero-mean Gaussian distributions, i.e., $\mathbf{x}_{\text{A}}, \mathbf{p}_{\text{A}} \sim \mathcal{N}\left(0_{N_{T}\times1}, V_s \undb{I}_{N_{T}}\right)$ with $V_s$ denoting the variance utilized for encoding the secret key information. Hence, the secret key is transmitted as the vector $\boldsymbol{\psi} = \mathbf{x}_{\text{A}} + \jmath \mathbf{p}_{\text{A}}$ from Alice's antenna elements. 

Recall that the estimation of the effective channel $\mathbf{H}_{\text{RIS}_{\text{LS}}}$, available to Alice and Bob, has been also made available to Eve via eavesdropping. Let $r$ and $\beta_{1},\ldots,\beta_{r}$ denote respectively this channel's rank and $r$ non-zero singular values. Applying SVD yields $\mathbf{H}_{\text{RIS}_{\text{LS}}} = \mathbf{U}_{\text{RIS}_{\text{LS}}} \mathbf{\Sigma_{\text{RIS}_{LS}}} \mathbf{V}_{\text{RIS}_{\text{LS}}}^\dagger$, where 
\beq
\mathbf{\Sigma}_{\text{RIS}_{\text{LS}}} \triangleq
\begin{bmatrix}
    \text{diag} \left(\sqrt{\beta_{1}},\ldots ,\sqrt{\beta_{r}}\right) & \! \! \! \boldsymbol{0}_{r \times\left(N_{T}-r\right)} \\
    \boldsymbol{0}_{\left(N_{R} -r\right) \times r} &\! \! \! \boldsymbol{0}_{\left(N_{R} -r\right) \times (N_{T}-r)}
\end{bmatrix}
\label{eq12}
\eeq 
with $\mathbf{V}_{\text{RIS}_{\text{LS}}}\in\mathbb{C}^{N_T\times N_T}$ and $\mathbf{U}_{\text{RIS}_{\text{LS}}}\in\mathbb{C}^{N_R\times N_R}$ being unitary matrices. This implies that Alice and Bob can respectively utilize $\mathbf{V}_{\text{RIS}_{\text{LS}}}$ and $\mathbf{U}_{\text{RIS}_{\text{LS}}}$ for transmit beamforming and receive combining, respectively. In addition, Eve can apply the latter matrices to $\mathbf{H}_{\text{RIS}_{\text{LS}}}$ to extract $\mathbf{\Sigma}_{\text{RIS}_{\text{LS}}}$ that carries the secret keys. To this end, Eve uses a collective entanglement attack \cite{2_ref_paper, neel_restricted_qkd} by creating a pair of two-mode squeezed vacuum (TMSV) modes $\{ e_{in_i}, e_{qm_i}\}$ $\forall i = 1,\ldots,N_{T} $ (also known as EPR pair), where $e_{in_i}$ is injected into Alice's transmitted signal and $e_{qm_i}$ is stored in the eavesdropper's quantum memory. Thus, the post-processed $N_R$-element received signal mode vector at Bob' side can be expressed as follows:
\beq
\mathbf{b} \triangleq \mathbf{U}_{\text{RIS}_{\text{LS}}}^\dagger\left( \mathbf{H}_{\text{RIS}} \mathbf{V}_{\text{RIS}_{\text{LS}}}\boldsymbol{\psi} + \mathbf{U}_{\text{RIS}_{\text{LS}}} \mathbf{S}_{\text{RIS}_{\text{LS}}} \mathbf{e}\right), 
\label{eq13}
\eeq
where $\boldsymbol{\psi} \triangleq \left[\psi_1, \ldots, \psi_{N_{T}}\right]^T$ denotes Alice's transmitting mode, $\mathbf{b} \triangleq \left[b_1, \ldots, b_{N_{R}}\right]^T$ is the received mode at Bob, and $\mathbf{e} \triangleq \left[e_1, \ldots, e_{N_{T}}\right]^T$ includes the Gaussian noise contributions injected by Eve to extract the secret key. In addition, the diagonal matrix $\textbf{S}_{\text{RIS}_{\text{LS}}}$ in \eqref{eq13} is given as:
\beq
\mathbf{S}_{\text{RIS}_{\text{LS}}} \triangleq
\begin{bmatrix}
    \text{diag} \left(\sqrt{1-\beta_{1}},\ldots ,\sqrt{1-\beta_{r}}\right) & \! \! \! \boldsymbol{0}_{r \times\left(N_{T}-r\right)} \\
    \boldsymbol{0}_{\left(N_{R} -r\right) \times r} &\mathbf{J}_{\left(N_{R} -r\right) \times \left(N_{T}-r\right)}
\end{bmatrix} \, , 
\label{eq14}
\eeq
where matrix $\mathbf{J}$ is defined as follows:
\beq
\mathbf{J} \triangleq
\begin{bmatrix}
    \textbf{I}_{N_1} & \! \! \! \boldsymbol{0}_{N \times \left(N_{T} -N_{R}\right)} \\
    \boldsymbol{0}_{\left(N_{R} -N_{T}\right) \times N} & \undb{I}_{N_2}
\end{bmatrix} \, ,
\label{eq14a}
\eeq
where $N = \min (\left(N_{R} -r\right),(N_{T}-r))$ and $N_1+N_2=N$.

By substituting expression (\ref{eq10}) for the estimated channel in (\ref{eq13}) and defining the channel estimation error matrix $\mathbf{\Delta H}_{\text{RIS}}\triangleq \mathbf{H}_{\text{RIS}_{\text{LS}}}-\mathbf{H}_{\text{RIS}}$, vector $\mathbf{b}$ can be re-expressed as:
\beq
\mathbf{b} = \mathbf{\Sigma_{\text{RIS}_{\text{LS}}} \boldsymbol{\psi}} - \underbrace{\mathbf{U}_{\text{RIS}_{\text{LS}}}^\dagger \mathbf{\Delta H}_{\text{RIS}} \mathbf{V}_{\text{RIS}_{\text{LS}}}\boldsymbol{\psi}}_{\triangleq\mathbf{n}_{{\text{RIS}}}} +  \mathbf{S}_{\text{RIS}_{\text{LS}}} \mathbf{e} \, , 
\label{eq15}
\eeq
where $\mathbf{n}_{\text{RIS}}\in\mathbb{C}^{N_R\times1}$ indicates the additional noise term that occurs as a result of the channel estimation errors.
 
Upon receiving $\undb{b}$, Bob conducts measurements to extract the confidential key information. It is to be recollected that during the channel estimation process, Bob utilized heterodyne measurements to estimate the effective channel $\mathbf{H}_{\text{RIS}}$, which entailed quantifying both the position and momentum quadratures of the incoming signal. However, during the key reception phase, we consider that Bob may perform either homodyne or heterodyne measurements. To this end, the measurement output of one or both real-valued quadratures can yield the secret key. Following the measurement, the correlation between Alice's input and Bob's output, delineated in terms of the quadratures, is computed $\forall i=1,2,\ldots,r$ as:
\beq
\hat{Q}_{b_i} \triangleq 
\sqrt{\beta_{i}} \hat{Q}_{a_i}+ \sqrt{1-\beta_{i}} \hat{Q}_{e_{i}} - n_{\text{RIS}_i} + n_{d_i},
\label{eq16}
\eeq
and subsequently, Eve's output mode can be expressed in terms of the quadratures $\forall i=1,2,\ldots,r$ as follows:
\beq
\hat{Q}_{{eo}_i} \triangleq 
- \sqrt{1-\beta_{i}} \hat{Q}_{a_i}+ \sqrt{\beta_{i}} \hat{Q}_{e_{i}} -n_{\text{RIS}_i},
\label{eq17}
 \eeq
where $\hat{Q}_{b_i}$ represents the measurement output of Bob's quadrature, $\hat{Q}_{a_i}$ denotes Alice's transmitted quadrature, and $\hat{Q}_{e_i}$ represents the quadrature of the mode injected by Eve to extract crucial information, all each $i$-th coherent state. In addition, all $\hat{Q}$'s in (\ref{eq16}) and (\ref{eq17}) denote one of the two quadratures, namely, either the position or momentum quadrature. Due to the presence of the preparation thermal noise, the variance of the quadrature corresponding to Alice's transmitted mode is given as $\text{var}(\hat{Q}_{a_i}) \triangleq V_a = V_s + V_o$. Similarly, the variance of the measurement occurring due to Eve's injected TMSV modes is defined $\text{var}(\hat{Q}_{e_{i}}) \triangleq V_e$. Owing to the statistics of the channel estimation noise, $n_{\text{RIS}_i}$ follows a zero-mean Gaussian distribution, implying that $n_{\text{RIS}_i} \sim \mathcal{N} \left(0, \sigma_{\text{RIS}_i}^2\right)$ with $\sigma_{\text{RIS}_i}^2 = 0.5 \mathbf{C_{\text{RIS}}} \left(i, i\right)$ where $\mathbf{C_{\text{RIS}}}$ represents the covariance matrix of the extra noise vector $\undb{n}_{\text{RIS}}$ in $(\ref{eq16})$. Additionally, $n_{d_i}$ denotes the detection noise arising during Bob's measurement, which also follows a zero-mean Gaussian distribution with a variance $\sigma_{d_i}^2 \triangleq d_m \left(\nu_{en} +1 \right) -1$, i.e., $n_{d_i} \sim \mathcal{N}\left(0, \sigma_{d_i}^2 \right)$, where $d_m$ takes the values $1$ or 2 for homodyne or heterodyne measurements, respectively.

The SKR of the system can be computed based on the available measurements in (\ref{eq16}). Hence, the secret key can be employed for key encryption and subsequent transmission, conditioned on the computed SKR being above a pre-decided threshold. However, for the SKR computation, the value of $\sigma_{\text{RIS}_i}^2$ must be ascertained to Alice and Bob, implying that the covariance matrix of $\undb{n}_{\text{RIS}}$ needs to be estimated. Using $(\ref{eq15})$ this covariance matrix, defined as $\mathbf{C_{\text{RIS}}}\in\mathbb{C}^{N_R\times N_R}$, can be expressed as follows:
\beqarr
\mathbf{C_{\text{RIS}}} \! \! \! \! &\triangleq& \! \! \! \!
\undb{E} \left[\mathbf{n}_{\text{RIS}} \mathbf{n}_{\text{RIS}}^\dagger\right] \nn \\
&=& \! \! \! \!
\undb{E} \left[ \textbf{U}_{\text{RIS}_{\text{LS}}}^\dagger \mathbf{\tilde N} \boldsymbol{\Psi}_p^\dagger
\left(\boldsymbol{\Psi}_p \boldsymbol{\Psi}_p^\dagger \right)^{-1}
\textbf{V}_{\text{RIS}_{\text{LS}}} \boldsymbol{\psi} \boldsymbol{\psi}^\dagger \right. \nn \\
&&\hspace{0.2cm} \left. \times
\textbf{V}_{\text{RIS}_{\text{LS}}}^\dagger
\left(\boldsymbol{\Psi}_p^\dagger \boldsymbol{\Psi}_p \right)^{-1} 
\boldsymbol{\Psi}_p \mathbf{\tilde N}^\dagger \textbf{U}_{\text{RIS}_{\text{LS}}} \right] \nn \\
&\stackrel{(a)}{=}& \! \! \! \!
2V_a \undb{E} \left[\textbf{U}_{\text{RIS}_{\text{LS}}}^\dagger
\mathbf{\tilde N} \boldsymbol{\Psi}_p^\dagger
\left(\boldsymbol{\Psi}_p \boldsymbol{\Psi}_p^\dagger \right)^{-1} 
\left(\boldsymbol{\Psi}_p^\dagger \boldsymbol{\Psi}_p \right)^{-1} 
\right. \nn \\ 
&& \quad\quad\hspace{0.05cm} \left. \times
\boldsymbol{\Psi}_p \mathbf{\tilde N}^\dagger \textbf{U}_{\text{RIS}_{\text{LS}}} \right] \nn \\
&\stackrel{(b)}{=}& \! \! \! \!
\frac{2V_a}{\left(V_p L_p \right)^2} \undb{E}\left[\textbf{U}_{\text{RIS}_{\text{LS}}}^\dagger \mathbf{\tilde N} \boldsymbol{\Psi}_p^\dagger
\boldsymbol{\Psi}_p \mathbf{\tilde N}^\dagger \textbf{U}_{\text{RIS}_{\text{LS}}}\right] \nn \\
&\stackrel{(c)}{=}& \! \! \! \!
\frac{2V_a N_{T}}
{V_pL_p} \textbf{U}_{\text{RIS}_{\text{LS}}}^\dagger \mathbf{C_{\tilde n}} \textbf{U}_{\text{RIS}_{\text{LS}}}\, ,
\label{eq18}
\eeqarr
where $\undb{E}\left[\boldsymbol{\psi} \boldsymbol{\psi}^\dagger\right] = 2V_a\textbf{I}_{N_{T}}$ was used in step $(a)$ and  $\boldsymbol{\Psi}_p \boldsymbol{\Psi}_p^\dagger = V_p L_p\textbf{I}_{N_{T}}$ was used in $(b)$. Furthermore, step $(c)$ follows from \cite[Lemma 4]{mckey_capacitybound_2005_1499046} and $\text{Tr}\left(\boldsymbol{\Psi}_p^\dagger
\boldsymbol{\Psi}_p\right) = V_p N_{T} L_p$.

Since the noise covariance matrix $\mathbf{C_{\tilde n}}=\undb{E}[\mathbf{\tilde N} \mathbf{\tilde N} ^\dagger ]$ in (\ref{eq18}) is unknown, the first step is to ascertain its maximum likelihood (ML) estimate. This estimate is subsequently utilized to obtain an estimate for $\mathbf{C_{\text{RIS}}}$. The ML estimate of $\mathbf{C_{\tilde n}}$ can be computed using the estimated channel matrix $\textbf{H}_{\text{RIS}_{\text{LS}}}$ and the known pilot symbol matrix $\boldsymbol{\Psi}_p$, as follows:
\beqarr
\mathbf{\hat{C}_{\tilde n_\text{ML}}} \! \! \! \! &\triangleq& \! \! \! \! 
\arg\max_{\mathbf{C_{\tilde n}}}
\log \left( \frac{1}{\pi^{N_{R}}\det \left(\mathbf{C_{\tilde n}} \right)} \right. \nn \\
&& \! \! \! \! \times
\exp \!\left\{ \!-\! \left(\mathbf{y}_p^{\ell} \!-\! \mathbf{H_{\text{RIS}_{LS}}} \boldsymbol{\psi}_p^\ell\right)^{\dagger} \! 
\mathbf{C^{-1}_{\tilde n}}
\!\!\left(\mathbf{y}_p^{\ell}\! -\! \mathbf{H_{\text{RIS}_{LS}}} \boldsymbol{\psi}_p^\ell\right) \!\!\right\} \Bigg) \nn \\
&=& \! \! \! \! \arg\min_{\mathbf{C_{\tilde n}}}
\log \left( \pi^{N_{R}}\det \left(\mathbf{C_{\tilde n}} \right) \right) \nn \\
&& \! \! \! \! + \sum_{\ell=1}^{L_p}
\left(\mathbf{y}_p^{\ell} \!-\! \mathbf{H_{\text{RIS}_{LS}}} \boldsymbol{\psi}_p^\ell\right)^{\dagger} \! 
\mathbf{C^{-1}_{\tilde n}}
\!\!\left(\mathbf{y}_p^{\ell}\! -\! \mathbf{H_{\text{RIS}_{LS}}} \boldsymbol{\psi}_p^\ell\right).
\label{eq19}
\eeqarr
Solving for $\mathbf{C_{\tilde n}}$ by computing (\ref{eq19})'s derivative, and then, equating it to zero results in the ML estimate of the covariance matrix, which is given by:
\beq
\mathbf{\hat{C}_{\tilde n_\text{ML}}} = \frac{1}{L_p} 
\sum_{t=1}^{L_p} \left( \mathbf{y}^{\ell} -  \mathbf{H_{\text{RIS}_{LS}}} \boldsymbol{\psi}_p^\ell \right) 
\left( \mathbf{y}^{\ell} -  \mathbf{H_{\text{RIS}_{LS}}} \boldsymbol{\psi}_p^\ell \right)^{\dagger}.
\label{eq20}
\eeq
Therefore, an estimate for $\mathbf{C_{\text{RIS}}}$ that can be used to determine the SKR is obtained by substituting $(\ref{eq20})$ in the place of $\mathbf{C_{\tilde n}}$ in~$(\ref{eq18})$, yielding:
\beq
\mathbf{\hat{C}_{\text{RIS}_\text{ML}}} = \frac{2V_a N_{T}}{V_p L_p} \textbf{U}_{\text{RIS}_{\text{LS}}}^\dagger \mathbf{\hat{C}_{\tilde n_{ML}}} \textbf{U}_{\text{RIS}_{\text{LS}}}.
\label{eq21}
\eeq
\section{SKR Analysis}
In this section, we analyze the SKR of the considered RIS-assisted MIMO CV-QKD THz system in the presence of channel estimation errors and the overheads derived in Section IV. 

Alice and Bob initiate the generation and reception of secret keys with the help of the two strings of correlated random quadrature (as mentioned after (\ref{eq17})) vectors, $\hat{\textbf{Q}}_{a_m}$ and $\hat{\textbf{Q}}_{b_m}$, with $m =1,\ldots, (T_c - L_p)$, where, $\hat{\textbf{Q}}_{a_m}$ represents the transmitted vectors of Gaussian-modulated coherent states by Alice and $\hat{\textbf{Q}}_{b_m}$ represents vectors of received quadrature states at Bob's end to employ detection technique, through the repetition of the QKD process outlined in Section IV, where $T_c$ represents the channel coherence time and $L_p$ denotes the pilot duration. The repeated execution of the protocol ensures that both gather enough data points to establish a statistically significant correlation needed for secure key generation.
To this end, they both perform a reconciliation or sifting protocol over a classical authenticated channel to retrieve the final keys, followed by error correction on the raw keys \cite{Weedbrook2018QMCrypto}. Typically, there exist two categories of reconciliation protocols. The first one is known as direct reconciliation (DR), in which Alice specifies the quadrature to be utilized for secret key generation, and the second one is termed reverse reconciliation (RR) according to which Bob indicates the quadrature to be employed for their measurement for secret key generation via a classical public channel. For the considered RIS-assisted MIMO CV-QKD system, it is expected that RR would result in higher SKR, since Eve who possesses the knowledge of the wireless channel, can extract more information when Alice specifies which quadrature is utilized for the secret key transmission. However, in RR, Bob's measured quadrature remains inaccessible to Eve. Therefore, we concentrate exclusively on RR, as this method can yield positive SKR for any channel transmittance $\beta_i \in [0,1]$. Conversely, for DR, it is necessary that $\beta_i > 0.5$ to attain positive SKRs \cite{ottaviani2020terahertz,Weedbrook2018QMCrypto}, which poses practical difficulties due to the considerably elevated path loss at THz frequencies, as given by (\ref{eq6}).

Following the secret key transmission, Eve uses a collective entanglement attack to maximize the extraction of key information, in which the received signals from Alice are measured one by one. Upon completion of Bob's measurements, Alice performs a collective measurement on the stored ancilla. To fix the error, Bob employs RR to achieve the positive secret key rate for any channel transmission. For such a setup, the SKR of each $i$-th ($i = 1,2,\ldots, r$) parallel channel is given as:
\begin{equation}
\begin{split}
\text{SKR}^{\text{RR}}_i \triangleq \left(1-\frac{L_p}{T_c} \right) 
\left( \eta I\left(Q_{a_i};Q_{b_i}\right) 
- I\left(Q_{b_i};E_i\right) \right),
\end{split}
\label{eq22}
\end{equation}
where $I\left(Q_{a_i}; Q_{b_i}\right)$ represents Shannon's mutual information between Alice's and Bob's measurement results. Similarly, $I\left(Q_{b_i}; E_i\right)$ denotes the quantum information (also known as Holevo information) between Eve's and Bob's quantum state for each $i$-th parallel channel. Furthermore, $\eta$ denotes the reconciliation efficiency and the term $\left(1-\frac{L_p}{T_c} \right)$ arises as a result of the channel estimation overhead. In addition, Shannon's mutual information between Alice's and Bob's measurement for each $i$-th parallel channel is computed as:
\beqarr
&& \! \! \! \! \! \! \! \! \! \! \!
\! \! \! \! \! \! \! \! \! \! \! \!
I\left(Q_{a_i};Q_{b_i}\right) \nn \\
&& \!\!\!\!\!\!\!  \!\!\!\!\!\! \!\!\!\!
\triangleq \frac{d_m}{2} \log_2 \left(1 + \frac{\beta_i V_s}
{\beta_i V_o +\left(1-\beta_i\right) V_e
+ \sigma_{d_i}^2 +\sigma_{\text{RIS}_i}^2} \right),
\label{eq23}
\eeqarr
where $d_m=1$ or $2$ depending on whether Bob performs homodyne or heterodyne measurements, respectively.

\subsection{The Holevo Information}
The Holevo information $I\left(Q_{b_i}; E_{i}\right)$, which represents the quantity of information exchanged between Eve and Bob's quantum state for each $i$-th parallel channel, is expressed as:
\beq
I\left(Q_{b_i};E_i\right) \triangleq S \left(E_i \right)
- S \left(E_i \left| Q_{b_i} \right. \! \right),
\label{eq24}
\eeq
where $S(E_i)$ and $S(E_i|Q_{b_i})$ denote respectively the Von Neumann entropy of Eve's output state and the conditional Von Neumann entropy of Eve's output state conditioned on Bob's quadrature mode $Q_{b_i}$. The latter is based on the choice of either homodyne or heterodyne measurements performed by Bob's side. For a given Gaussian quantum state $\tilde{\rho}$ (either $E_i$ or $E_i$ conditioned on $Q_{b_i}$), the Von Neumann entropy can be expressed as follows~\cite{weedbrook_gaoussian_van_numen}:
\beq
S\left( \tilde{\rho} \right) \triangleq \sum_{q=1}^2 h_o \left( \lambda_q \right),
\label{eq25}
\eeq
where $\lambda_q$'s are the symplectic eigenvalues of the covariance matrices of the Gaussian quantum states and the function $h_o (\cdot)$ outputs the Holevo information for $q=1$ and $2$:
\beqarr
h_o (\lambda_q ) \! \! \! \! &\triangleq& \! \! \! \! \left(\frac{\lambda_q+1}{2}\right)
\log_2 \left( \frac{\lambda_q+1}{2} \right) \nn \\
&& - \left( \frac{\lambda_q-1}{2} \right)
\log_2 \left( \frac{\lambda_q-1}{2} \right).
\label{eq26}
\eeqarr
It is noted that $q$ takes two values which correspond to the eigenvalues of the two covariance matrices corresponding to the quantum states. The first covariance matrix is the covariance matrix of Eve's state, which, for each $i$-th channel is expressed as follows:
\beq
\mathbf{\Sigma}_{E_i} =
\begin{bmatrix}
    V_{{eo}_i} \textbf{I}_2 & V_{c_i}\textbf{Z}\\
    V_{c_i}^* \textbf{Z}^T & V_e \textbf{I}_2
\end{bmatrix},
\label{eq28}
\eeq
where $V_{{eo}_i} \triangleq \left( 1-\beta_i \right)V_a + \beta_i V_e + \sigma_{\text{RIS}_i}^2$ and $V_{c_i} \triangleq \sqrt{\beta_i \left(V_e^2 -1 \right)}$. Recall that $V_e$ is the variance of Eve's generated quantum state. Furthermore, $\textbf{Z}$ is the Pauli-z matrix expressed as $\text{diag}\left(1,-1  \right)$. Thus, the symplectic eigenvalues of $\mathbf{\Sigma}_{E_i}$ are computed using the results in~\cite{2_ref_paper, Weedbrook2018QMCrypto} as:
\beq
\lambda_{i_{1,2}} =\sqrt{\frac{1}{2}\left( \nabla_{i}
\pm \sqrt{\nabla_{i}^2
- 4\text{det} \left(\mathbf{\Sigma}_{E_i} \right)}\right)} \, ,
\label{eq29}
\eeq
where we have used the definitions:
\begin{align}
\label{eq30}{\nabla}_{i} &\triangleq \left(1-\beta_i\right)^2\left(V_a^2 + V_e^2\right) + 2\beta_i\left(1-\beta_i\right)V_aV_e \\& 
\quad\,\,
+ 2\beta_i+ 2\left(\left(1-\beta_i\right)V_a + \beta_i V_e  \right)\sigma_{\text{RIS}_i}^2 + \sigma_{\text{RIS}_i}^4,\nonumber\\
\label{eq31}\text{det} \left(\mathbf{\Sigma}_{E_i} \right) &\triangleq \left(\left(1-\beta_i\right)V_a V_e + \sigma_{\text{RIS}_i}^2 V_e + \beta_i  \right)^2.
\end{align}
Thus, the Von Neumann entropy of Eve's state is obtained as:
\beq
S \left(E_i \right) = h_o(\lambda_{i_1}) + h_o(\lambda_{i_2}) \, .
\label{eq32}
\eeq

The second covariance matrix $\mathbf{\Sigma}_{E_{i}|Q_{b_i}}$ corresponds to Eve's conditional state given Bob's measurement. Thus, when Bob performs a homodyne or heterodyne measurement, this matrix can be determined using general Gaussian measurements, and it is thus given for each $i$-th channel as follows:
\beq
\mathbf{\Sigma}_{E_{i}|Q_{b_i}} = \mathbf{\Sigma}_{E_{i}} - \frac{1}{V_{b_i}}\textbf{W}_{i}\mathbf{\Pi} \textbf{W}_{i}^\dagger \, ,
\label{eq33}
\eeq
where we have used the matrix definitions:
\bsub
\beq
\mathbf{\Pi} \triangleq\begin{cases}
    \text{diag}(1,0),\quad \text{homodyne}  \\
    \text{diag}(1,1),\quad \text{heterodyne} 
\end{cases}\!\!\!\!\!,
\label{eq34}
\eeq
\beq
\textbf{W}_{i} \triangleq \begin{bmatrix}
     \xi_i \textbf{I}_2 \\
     \varsigma_i \textbf{Z}
 \end{bmatrix} ,
\label{eq35}
\eeq
as well as the parameter:
\beq
V_{b_i} =\begin{cases}
     \beta_i V_o+ \left(1-\beta_i\right) V_e + \nu_{en}+\sigma_{\text{RIS}_i}^2, \text{homodyne} \\
     \beta_i V_o + \left(1-\beta_i\right) V_e + 2\nu_{en}+1 +\sigma_{\text{RIS}_i}^2, \text{heterodyne}
\end{cases}\!\!\!\!\!,
\label{eq36}
\eeq
\esub
Substituting (\ref{eq34}), (\ref{eq35}), and (\ref{eq36}) in (\ref{eq33}) followed by algebraic simplifications results in the following expression:
\beq
\mathbf{\Sigma}_{E_{i}|Q_{b_i}} =
\begin{bmatrix}
   \textbf{A}_{i} & \textbf{C}_{i} \\
    \textbf{C}^\dagger_{i} & \textbf{B}_{i}
\end{bmatrix},
\label{eq37}
\eeq
which is of the same form as (\ref{eq28}), and thus, its symplectic eigenvalues can be computed as follows:
\beq
\lambda_{i_{3,4}} =\sqrt{\frac{1}{2}\left( \tilde{\nabla}_{i}
\pm \sqrt{\tilde{\nabla}_{i}^2
- 4\text{det} \left(\mathbf{\Sigma}_{E_{i}|Q_{b_i}} \right)}\right)} \, ,
\label{eq38}
\eeq
where we used the definitions:
\begin{align}
\tilde{\nabla}_{i} &\triangleq \text{det} \left( \textbf{A}_{i} \right)
+ \text{det} \left( \textbf{B}_{i} \right)
+ 2  \text{det} \left( \textbf{C}_{i} \right), \\ 
\textbf{A}_{i} &\triangleq
\text{diag} \left( \left( V_{eo_i} -\frac{\xi_i^2}{V_{b_i}}\right),
V_{eo_i}\right),\\
\textbf{B}_{i} &\triangleq
\text{diag} \left( \left(V_{e} -\frac{\varsigma_i^2}{V_{b_i}} \right), V_{e}\right),\\
\textbf{C}_{i} &\triangleq
\text{diag} \left( \left(V_{c_i} -\frac{\xi_i \varsigma_i}{V_{b_i}}\right) ,- V_{c_i}\right),\\
\xi_i &\triangleq
\sqrt{\beta_i\left(  1-\beta_i\right)}\left(V_e -V_a  \right)
+ \sigma_{\text{RIS}_i}^2,\\
\varsigma_i &\triangleq
\sqrt{\left(1-\beta_i  \right)\left(V_e^2 -1  \right)}.
\end{align}
Hence, the conditional Von Neumann entropy of Eve's state given Bob's measurement is given by:
\beq
S \left(E_i \left| Q_{b_i} \right. \!\right) = h_o(\lambda_{i_3}) + h_o(\lambda_{i_4})
\label{eq42}
\eeq
with $h_o(\cdot)$ given in (\ref{eq26}), and $\lambda_{i_3}$ and $\lambda_{i_4}$ given in (\ref{eq38}). 

\subsection{Closed-Form SKR Formula}
By using (\ref{eq22}), (\ref{eq23}), and (\ref{eq24}), the SKR of the considered RIS-assisted MIMO CV-QKD system employing collective entanglement attack and RR at Bob's side is obtained as:
\beqarr
&& \! \! \! \! \! \! \! \! \! \! \! \! \! \! \!
\text{SKR}^{\text{RR}}_\text{MIMO}
= \left(1-\frac{L_p}{T_c} \right) \nn \\
&& \! \! \! \! \! \! \! \! \! \! \! \! \! \!
\times \sum_{i=1}^r \left( \frac{\eta d_m}{2}
\log_2 \left( 1 + \frac{\beta_i V_s}
{\beta_i V_o +\left(1-\beta_i\right) V_e + \sigma_{d}^2 
+\sigma_{\text{RIS}_i}^2} \right) \right. \nn \\ 
&& \quad
- h_o(\lambda_{i_1}) - h_o(\lambda_{i_2})
+ h_o(\lambda_{i_3}) + h_o(\lambda_{i_4}) \bigg) .
\label{eq43}
\eeqarr
\section{Numerical Results and Discussion}
This section presents numerical results validating the analytical framework outlined in the previous sections. The simulation studies were carried out considering standard system parameters. The THz frequency was set as $f_c=15$ THz considering $T_e = 297$ K, 
and the atmospheric absorption coefficient was chosen as $\rho = 50$ dB/km. The antenna gain was set as $G_a = 30$ dBi, the pilot power was $V_p = 60$ dB, and the variance of Alice's modulated signal was chosen as $V_a = V_s +V_o$ with $V_s = 1$ and $V_o$ as discussed in Section~III (i.e., $V_o = 2 \Bar{n} + 1$ with $\Bar{n} = [\exp \left(hf_c/k_BT_e \right) - 1]^{-1}$). 
We have assumed that all $K$ unit cells of the RIS share the same phase configuration, i.e., $\phi_k =\phi$ $\forall k \in \{1,\ldots, K\}$. 

\begin{figure}[!t]
    \centering
    \includegraphics[width= 9cm, height=6cm]{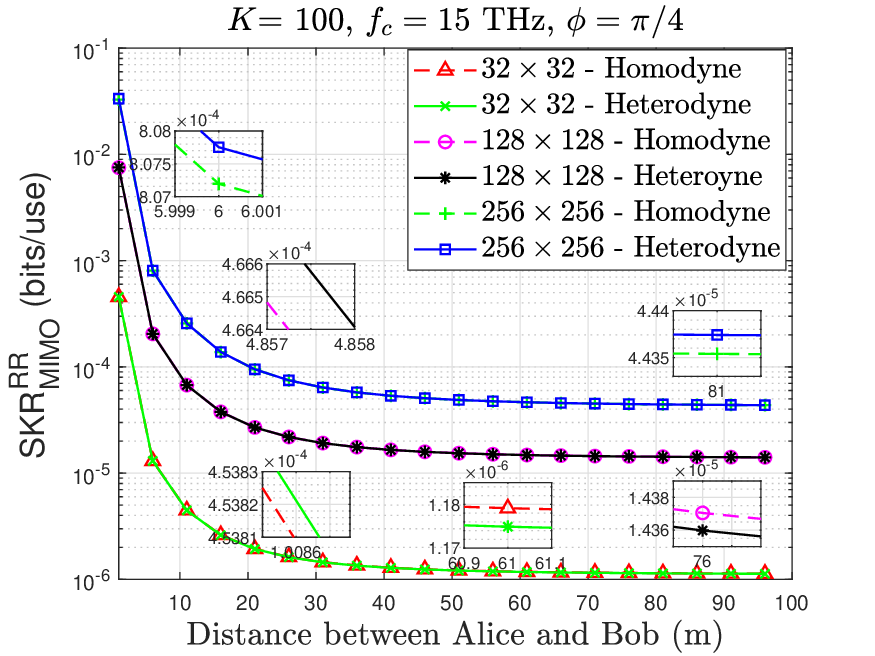}
    \caption{SKR versus the transmission distance between Alice and Bob for $N_{T}= N_{R}=\{32, 128, 256\}$, $f_c = 15$ THz, $K = 100$, and $\phi = \pi/4$, with Bob employing homodyne and heterodyne measurements for secret key detection.}
    \label{f2}
\end{figure}
Fig.~\ref{f2} illustrates the SKR (in bits/channel use) versus the transmission distance $d$ (in meters) between Alice and Bob for various MIMO configurations considering $K=100$ RIS unit elements and reconciliation efficiency $\eta = 0.95$. The plots were obtained using (\ref{eq43}) and the values of $d_m$, $\sigma_{d_i}^2$, and $\lambda_{i_t} \forall t \in \{1,\ldots,4\}$ depended on the choice for homodyne and heterodyne measurements, as given in (\ref{eq35}) and (\ref{eq37}). The ML estimate $\mathbf{\hat{C}_{\tilde n_{ML}}}$ obtained in (\ref{eq21}) was used in (\ref{eq22}) for the evaluation of the noise variance arising due to the channel estimation error $\sigma_{\text{RIS}_i}^2$. It can observed from the figure that, in the $256 \times 256$ MIMO configuration, the SKR obtained by employing heterodyne detection is higher than that obtained when Bob employs homodyne measurements. This trend is consistent for all distance values between Alice and Bob. However, for lower MIMO configurations, such as the considered $128 \times 128$ and $32 \times 32$ case, it is observed that heterodyne measurements result in a higher SKR value at lower distance values; this trend becomes reversed at higher distances. This behavior can be justified from the fact that, for lower MIMO configurations, the noise caused by heterodyne detection (due to the exposure of both the position and momentum quadrature) is greater than that of homodyne detection at higher distances between the transceiver pair. In addition, the SKR of the system increases when $N_T$ and $N_R$ are increased. It can also be seen that SKR saturates at higher distance values, which can be attributed to the use of the RIS, without which the SKR would reduce with increasing distance between Alice and Bob.
\begin{figure}[!t]
    \centering
    \includegraphics[width= 9cm, height=6cm]{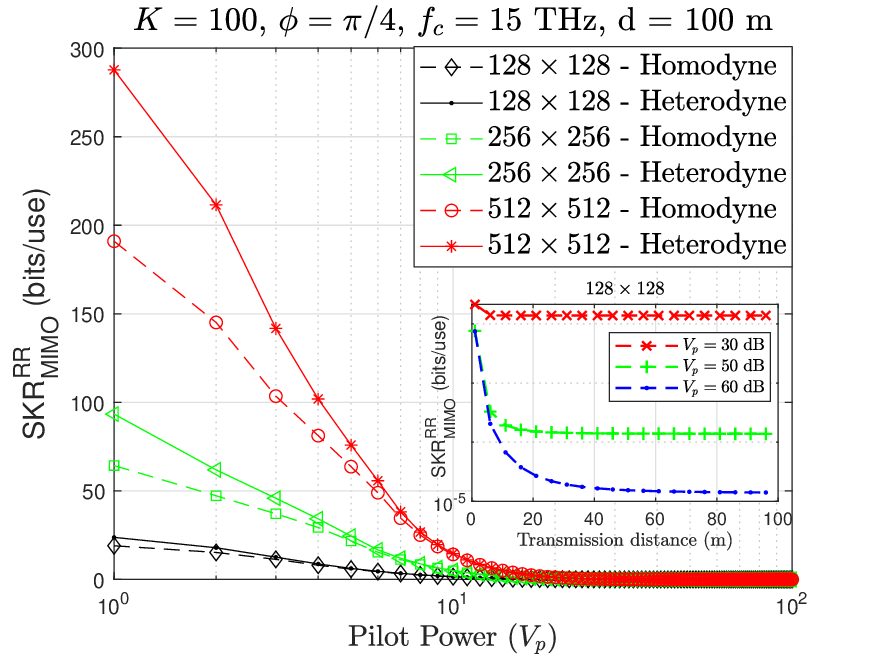}
    \caption{SKR versus the pilot power $V_p$ for $N_{T}= N_{R}= \{128, 256, 512\}$ MIMO configurations, $f_c = 15$~THz, $d=100$~m, $K = 100$, and $\phi = \pi/4$, with Bob employing homodyne and heterodyne measurements for secret key detection.}
    \label{f3}
\end{figure}
The SKR as a function of the pilot power $V_p$ for different RIS-assisted MIMO configurations at a constant transmission distance of $d = 100$ m, with Bob conducting either homodyne or heterodyne measurements, is depicted in Fig.~\ref{f3}. As shown, the performance of heterodyne detection surpasses that of homodyne detection at lower $V_p$ values, and the gap between the SKRs arising due to these two measurements is more prominent at higher values of $N_T$ and $N_R$. It can be also observed that the SKR degrades with increasing $V_p$. This can be intuitively understood from the fact that a higher pilot power implies a longer duration for channel estimation, resulting in a lower duration for secret key transmissions, thus, leading to SKR values. This effect is also mathematically evident from expression~(\ref{eq20}) where the duration of the pilot symbol occurs both in the summation term and in the denominator, thus, nullifying the effect of the increased pilot power in the estimation of the noise variance. As seen in the figure, this effect is valid for all $d$ values. There is a also saturation effect becoming prominent at lower transmission distances for lower pilot power values.

\begin{figure}[!t]
    \centering
    \includegraphics[width= 9cm, height=6cm]{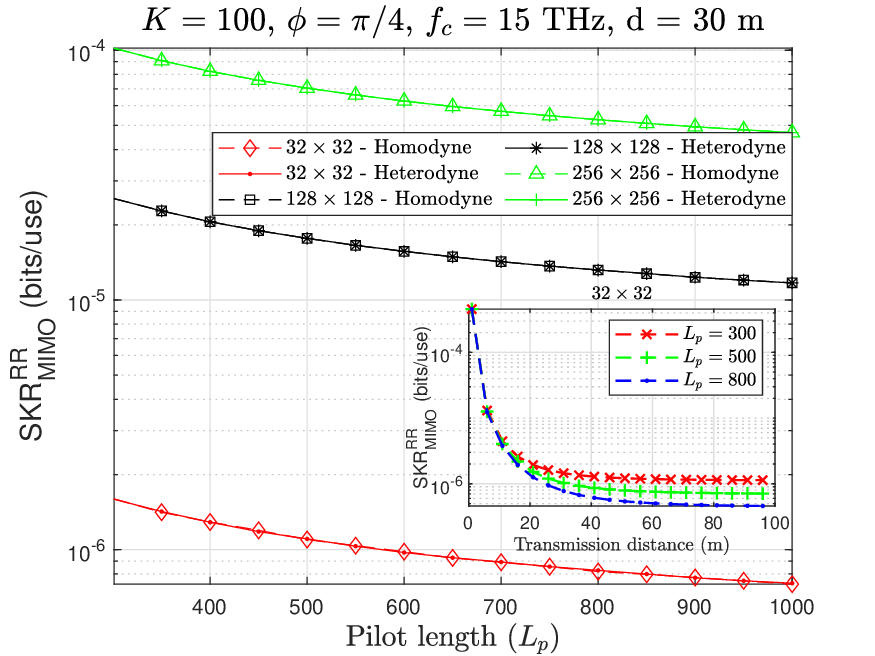}
    \caption{SKR versus the pilot length $L_p$ for $N_{T}= N_{R}= \{32, 128, 256\}$ MIMO configurations, $f_c = 15$~THz, $d=30$~m, $K = 100$, $\phi = \pi/4$, with Bob employing homodyne and heterodyne measurements for secret key detection.}
    \label{f4}
\end{figure}
\begin{figure}[!t]
    \centering
    \includegraphics[width= 9cm, height=6cm]{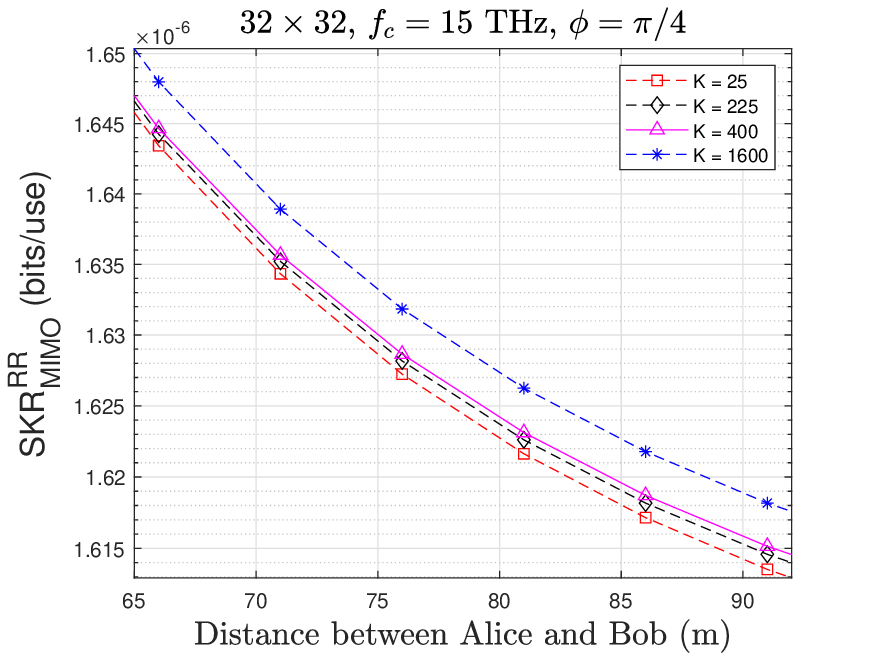}
    \caption{SKR versus transmission distance $d$ between Alice and Bob for a $32 \times 32$ MIMO configuration, $f_c=15$~THz, $\phi = \pi/4$ for $K = \{25, 225, 400, 1600\}$ RIS elements, with homodyne detection at Bob's side.}
    \label{f5}
\end{figure}

\begin{figure*}[!t]
     \centering
     \begin{subfigure}[b]{0.5\textwidth}
         \centering
        \includegraphics[width=\textwidth,height=10cm]{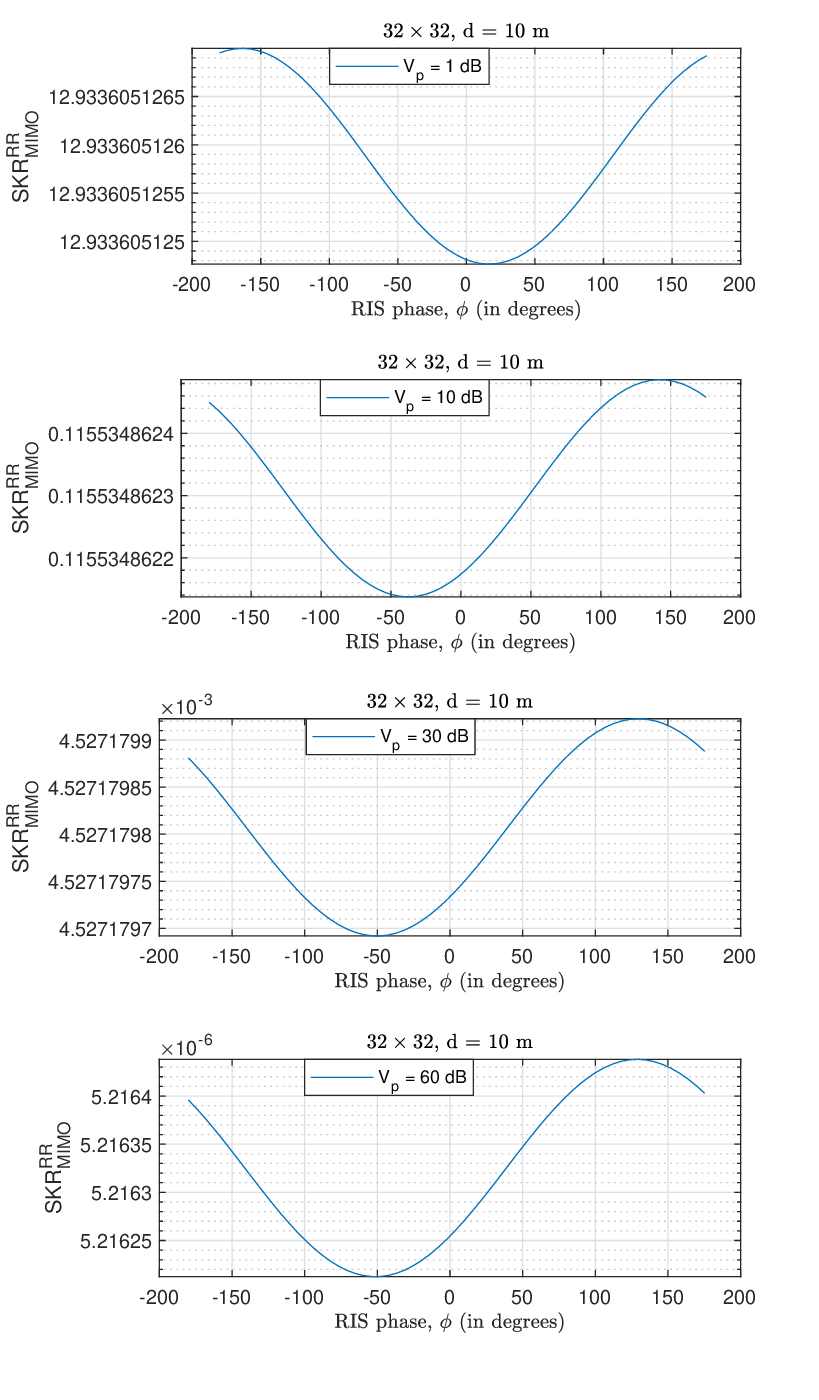}
         \caption{$32 \times 32$ at $d = 10$ m}
         \label{fig;6a}
     \end{subfigure}
     \hspace{-0.3cm}
     \begin{subfigure}[b]{0.5\textwidth}
         \centering
        \includegraphics[width=\textwidth,height=10cm]{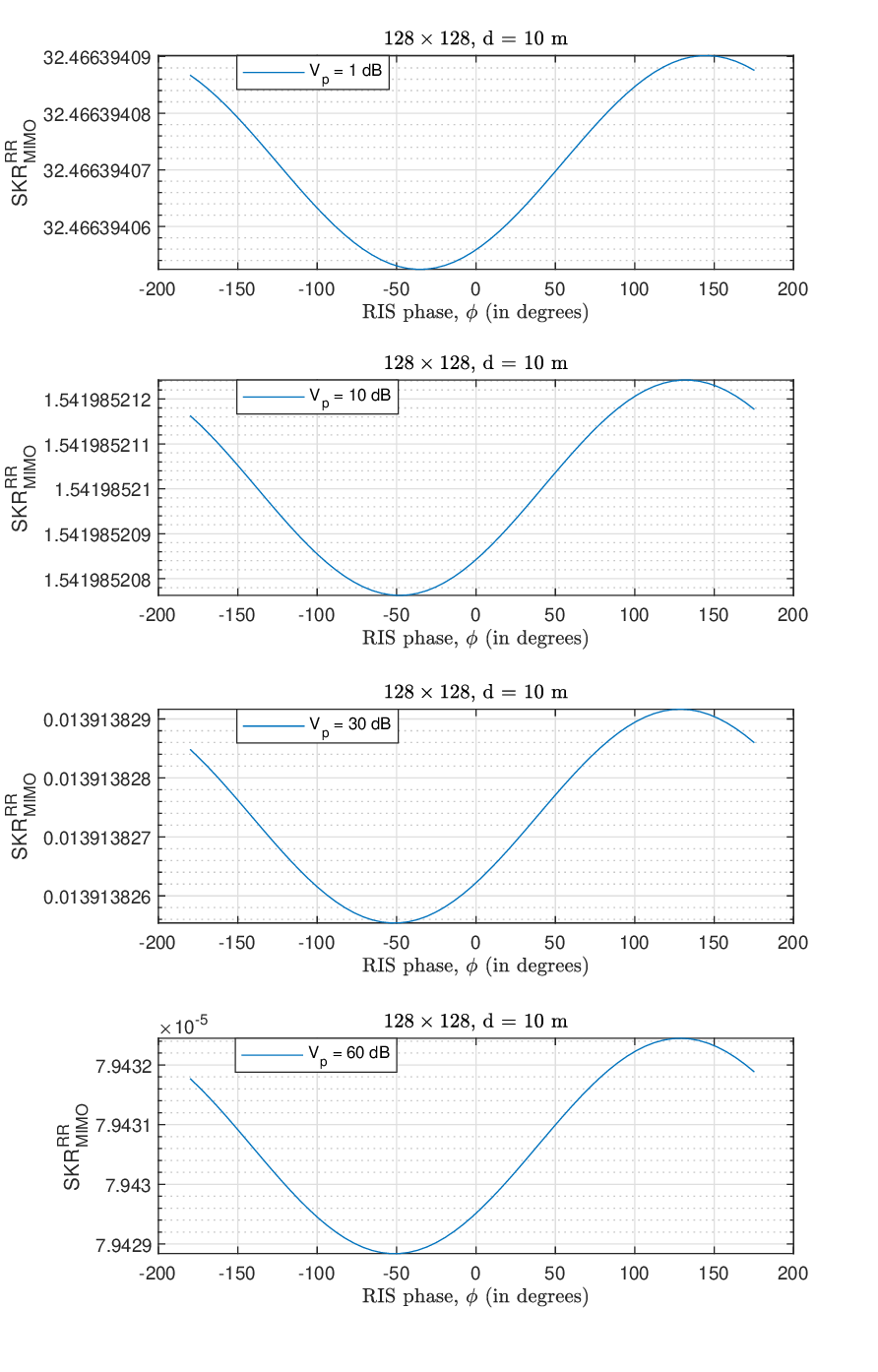}
         \caption{$128 \times 128$ at $d = 10$ m }
         \label{fig;6b}
     \end{subfigure}
     \hspace{-0.3cm}\\
     \begin{subfigure}[b]{0.5\textwidth}
         \centering
        \includegraphics[width=\textwidth,height=10cm]{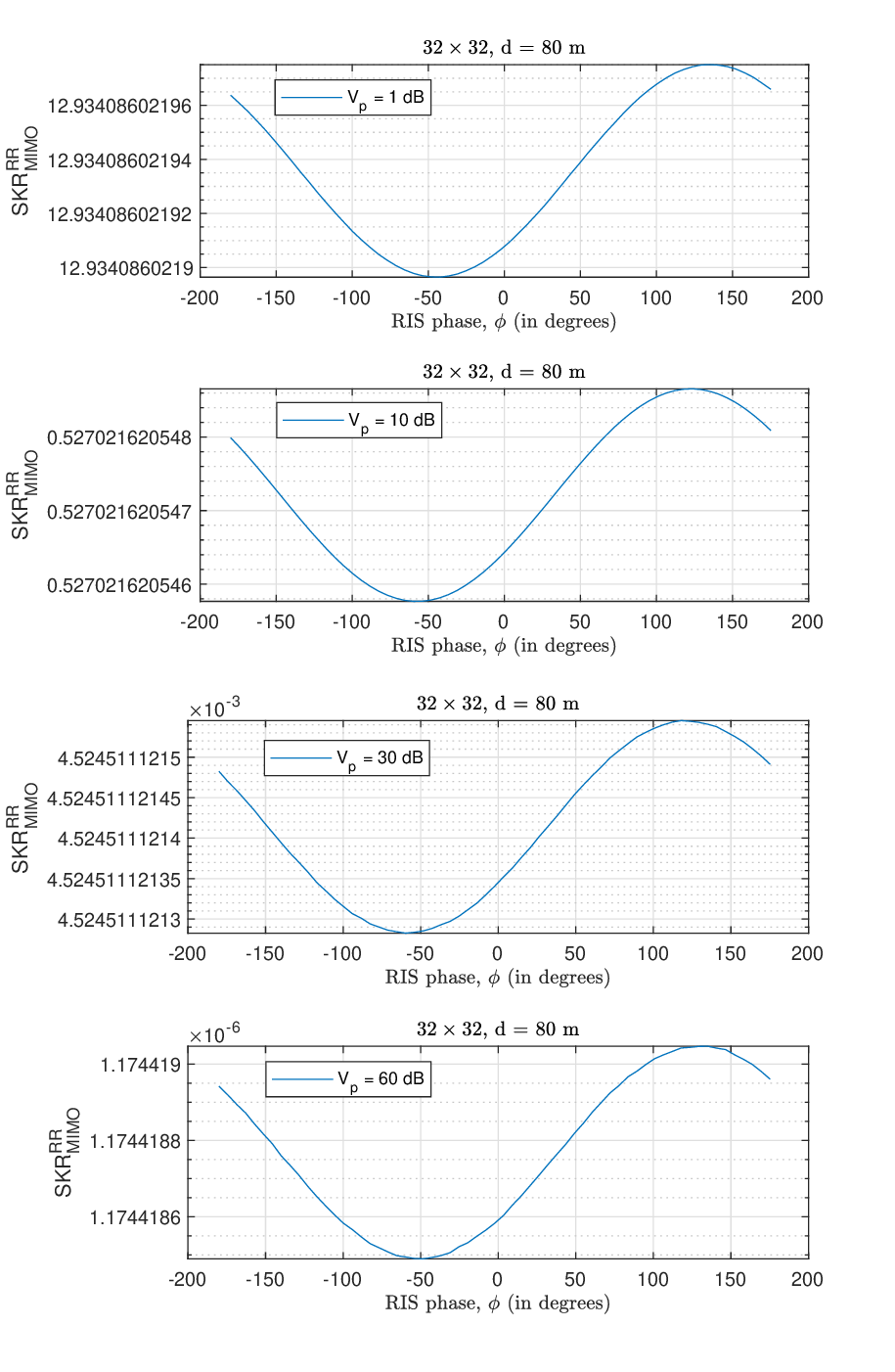}
         \caption{$32 \times 32$ at $d = 80$ m}
         \label{fig;6c}
     \end{subfigure}
          \hspace{-0.3cm}
     \begin{subfigure}[b]{0.5\textwidth}
         \centering
        \includegraphics[width=\textwidth,height=10cm]{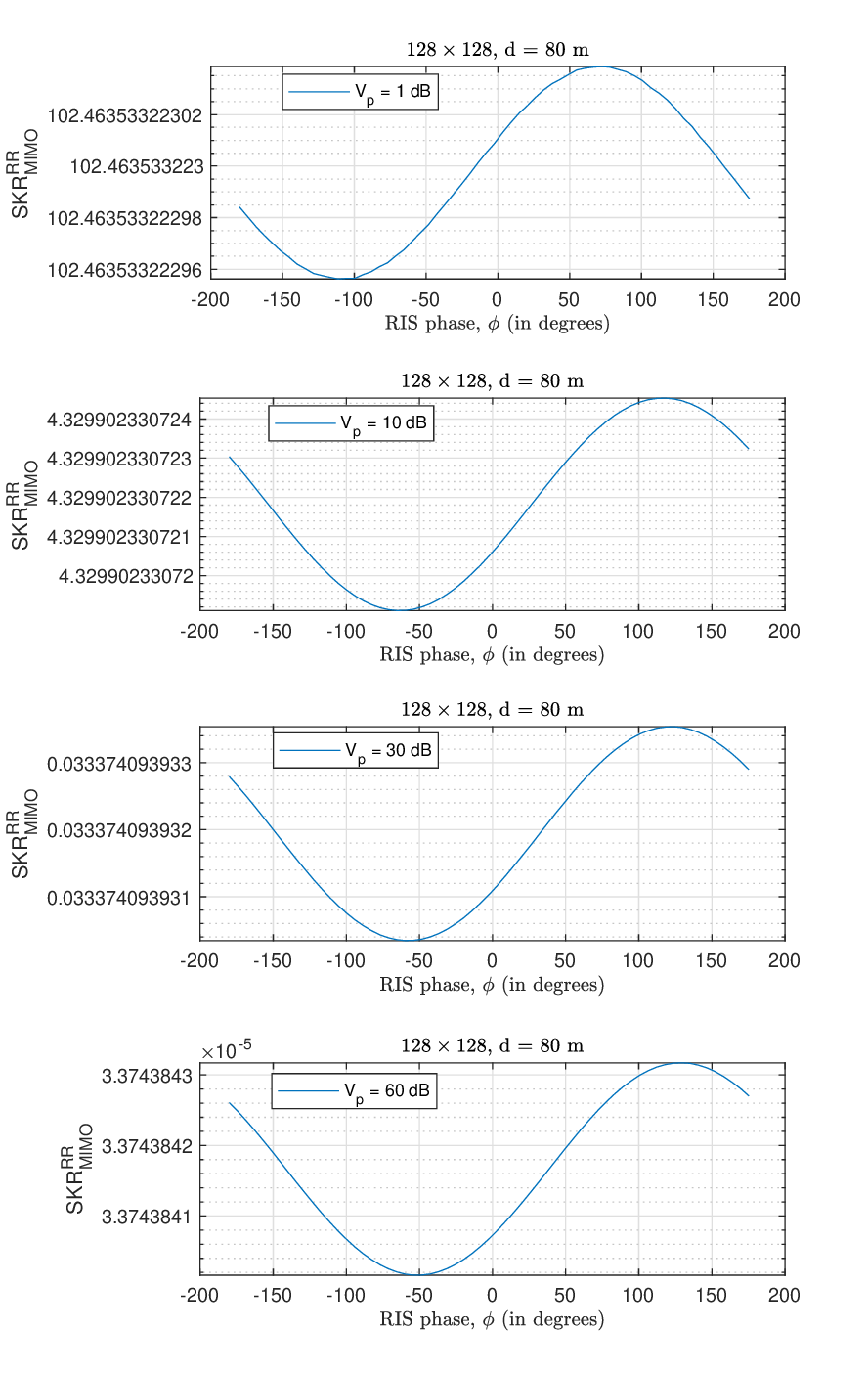}
         \caption{$128 \times 128$ at $d = 80$ m }
         \label{fig;6d}
     \end{subfigure}
        \caption{SKR versus the RIS phase configuration $\phi$ for $N_{T} = N_{R} =\{32, 12\}$ MIMO configurations, $f_c=15$~THz, $d=\{10, 80\}$~m, and $K = 100$, with Bob employing heterodyne detection for secret key detection.}
        \label{f6}
\end{figure*}
\begin{table*}[!t]
\caption{Optimal phase value for all RIS elements for $N_{T}=N_{R} =\{16,32, 64, 128\}$, pilot powers $V_p =\{1, 10, 30 ,60\}$ dB, and distances between Alice and Bob $d =\{10, 50, 80\}$ m.}
\centering
\renewcommand{\arraystretch}{2}
\resizebox{\textwidth}{!}{ 
\begin{tabular}{|c|c|c|c|c|c|c|c|c|c|c|c|c|c|c|}
\hline
\multirow{3}{*}{$N_{T} = N_{R}$} & \multicolumn{12}{c|}{Optimal phase shift, $\phi$, of RIS elements} \\ \cline{2-13} {}
 & \multicolumn{4}{c|}{ $d = 10$ m} & \multicolumn{4}{c|}{ $d = 50$ m} &\multicolumn{4}{c|}{ $d = 80$ m} \\ \cline{2-13}
 & $V_p = 1$ dB & $V_p = 10$ dB & $V_p = 30$ dB & $V_p = 60$ dB 
 & $V_p = 1$ dB & $V_p = 10$ dB & $V_p = 30$ dB & $V_p = 60$ dB 
 & $V_p = 1$ dB & $V_p = 10$ dB & $V_p = 30$ dB & $V_p = 60$ dB \\ \hline
16  & $-128.43^{\circ}$ & $72.10^{\circ}$  & $158.05^{\circ}$ & $129.40^{\circ}$  & $-151.35^{\circ}$ & $-99.78^{\circ} $& $112.20^{\circ}$ & $129.40^{\circ}$ 
& $43.45^{\circ}$  & $100.75^{\circ}$  & $-2.38^{\circ}$ & $129.40^{\circ}$\\ \hline
32  & $-162.81^{\circ}$ & $140.86^{\circ} $    & $129.40^{\circ}$ & $129.40^{\circ}$ 
& $152.32^{\circ}$  & $77.83^{\circ}$  & $129.40^{\circ}$ & $129.40^{\circ}$
& $135.13^{\circ}$  & $123.67^{\circ}$  & $117.94^{\circ}$ & $129.40^{\circ}$ \\ \hline
64  & $-180^{\circ} $ & $-105.52^{\circ}$ & $-134.16 ^{\circ}$ & $129.40^{\circ} $ & $-139.89^{\circ}$ & $72.10^{\circ}$  & $135.13^{\circ}$ & $129.40^{\circ}$ 
& $-162.81^{\circ}$  & $-162.81^{\circ}$  & $112.21^{\circ}$ & $129.40^{\circ}$ \\ \hline
128 & $146.59^{\circ}$  & $129.40^{\circ}$ & $129.40^{\circ}$ & $129.40^{\circ}$  & $-180^{\circ}$   & $-162.81^{\circ}$ & $129.40^{\circ}$ & $ 129.40^{\circ}$ 
& $102.46^{\circ}$  & $117.94^{\circ}$  & $123.66^{\circ}$ & $129.40^{\circ}$ \\ \hline
\end{tabular}
}
\label{t1}
\end{table*}
The impact of pilot duration on the SKR performance of the considered RIS-assisted MIMO CV-QKD system is showcased in Fig.~\ref{f4}, which demonstrates the SKR for both homodyne and heterodyne measurements conducted by Bob for varying values of $N_T$ and $N_R$ at a constant transmission distance of $d=30$ m between Alice and Bob. Similar to Fig.~\ref{f3}, SKR degrades as $L_p$ increases. This is indicative of the fact that the impact of pilot length is greater on the noise covariance, as demonstrated in (\ref{eq20}). Therein, the summation over the $L_p$ term predominates the $1/L_p$ component (see  (\ref{eq20}) and (\ref{eq21})). However, contrary to the observation in Fig.~\ref{f3}, the values of the SKR do not merge irrespective of the chosen MIMO configuration. Furthermore, there is a negligible difference in the performance of the homodyne and heterodyne detectors for varying $L_p$. Additionally, the saturation effect is not as prominent with increasing $L_p$ as it was with increasing $V_p$. A similar trend to $V_p$ is also observed for different values of $L_p$ at $d$ values. However, in contrast to Fig.~\ref{f3}, the effect of $L_p$ on the SKR is more prominent at higher values of $d$.

The advantage of employing an RIS to achieve a higher SKR is illustrated in Fig.~\ref{f5}, which depicts SKR with respect to different $d$ and $K$ values for a $32 \times 32$ MIMO configuration and $\phi = \pi/4$. It is demonstrated that the SKR of the system improves with increasing $K$, and this improvement becomes slightly larger for larger $d$ values. However, it is observed that the difference between the curves is small as compared to the order in which the number of elements in the RIS is increased. Thus, a small RIS becomes practical to be deployable for such systems.

Finally, the effect of the variation of the common phase shift $\phi$ of all RIS elements on the system's SKR performance is demonstrated in Fig.~\ref{f6}. It is noted that, to the best of our knowledge, this is the first work proposing the use of RISs for such quantum communication systems employing a practical channel estimation scheme, thus, it is our goal to show the SKR behaviour for various $\phi$ values. As shown, SKR depends on $\phi$ in a sinusoidal manner. This implies that there exists an optimal $\phi$ maximizing the SKR performance. Table~\ref{t1} includes such values for some of the system parameters. It can be seen that both $V_p$ and $d$ values play a crucial role in the optimal $\phi$ maximizing SKR. For a given MIMO configuration, increasing the value of $V_p$ at a given distance, initially reduces the value of the optimal $\phi$, which then tends to saturate at higher values of $V_p$. However, there is no such straightforward trend for the relation of the optimal $\phi$ with value of $d$, which can be attributed to the highly non-linear nature of the THz channel model. Similarly holds for the relation between the optimal $\phi$ and the $N_T=N_R$ values.
\section{Conclusions}
This paper presented an RIS-assisted point-to-point MIMO communication system operating at THz frequencies where the transmitter, Alice, employs CV-QKD to transmit secret keys to the receiver, Bob. We considered the practical case where Bob performs LS-based channel estimation and feeds back the MIMO channel coefficients to Alice, prior to receiving secret keys from her. An eavesdropper, Eve, was considered to intercept the secret key transmission, eavesdropping first the feedback channel with the channel estimation information, and then, during the secret key transmission, applying a collective Gaussian attack. Considering that Bob utilizes either homodyne or heterodyne detection for the reception of the secret key, we presented a novel expression for the SKR performance of the system. Extensive numerical results were presented to corroborate the analytical framework of the paper and indicate the variation of the system performance over different parameters. It was showcased that an RIS plays a crucial role in improving SKR, preventing this metric's degradation at higher link distances, and that there exists an optimal RIS phase configuration that boosts SKR even further. This optimal configuration was shown numerically to depend on the pilot signal power and the link distance. It was also demonstrated that heterodyne detection is preferable over homodyne at lower pilot symbol powers, providing additional design benefits for the RIS-assisted MIMO CV-QKD communication systems at the THz frequency band.
\bibliographystyle{IEEEtran}
\bibliography{IEEEabrv,refs,bibliography}
\end{document}